\documentclass[a4paper]{article}

\usepackage{18lomcon}        
\usepackage{cite}             
\usepackage{epsfig}           
\usepackage{epstopdf}

\begin{document}


\title{NEUTRINO PHYSICS AND LEPTONIC WEAK BASIS INVARIANTS}

\author{M. N. Rebelo \email{rebelo@tecnico.ulisboa.pt \\
Invited talk given at 18th Lomonosov Conference on Elementary Particle Physics (Moscow, Russia, August 24-30, 2017)}
}

\affiliation{Centro de F\'isica Te\'orica de Part\'iculas -- CFTP and Dept de F\' \i sica,
Instituto Superior T\'ecnico -- IST, Universidade de Lisboa (UL), Av. Rovisco Pais,
P-1049-001 Lisboa, Portugal}


\date{}
\maketitle


\begin{abstract}
   In this talk we present a powerful tool applied to the study of Leptonic Physics.
This tool is based on the construction of Weak Basis invariant relations
associated to different properties of leptonic models. The rationale behind
these constructions is the fact that fermion mass matrices related through weak basis 
transformations look different but lead to the same physics. Such invariants can be 
built, for instance, with the aim to test leptonic models for different types of CP 
violation. These invariants are also relevant beyond such tests and have been applied 
to the study of implications from zero textures appearing in the leptonic mass matrices. 
In this case an important question is, how can a flavour model corresponding to a 
set of texture zeros be recognised, when written in a different weak basis, where the zeros are 
not explicitly present. Another important application is the construction of invariants 
sensitive to the neutrino mass ordering and the $\theta_{23}$ octant.
\end{abstract}

\section{Introduction}

One of the major puzzles in Particle Physics is the origin of fermion masses, 
mixing and CP violation, the so-called Flavour Puzzle. It is by now established
that at least two of the three active light neutrinos have non-zero masses. 
In the Standard Model neutrinos are strictly massless. Accounting for neutrino
masses requires physics beyond the Standard Model and has profound phenomenological 
implications. Neutrino Physics is at present an important field of research
with many different experiments taking data and future new facilities and
upgrades being planned. Among the fundamental open questions in this field are
whether or not neutrinos are Dirac or Majorana particles, whether or not there is 
CP violation in the leptonic sector, what is the absolute neutrino mass scale and
what is the mass ordering, meaning, what is the sign of $m^2_3 - m^2_1$? Is
the leptonic mixing matrix, $U_{PMNS}$ unitary? Are there sterile neutrinos? Do
neutrinos have nonstandard interactions?

Attempts at solving the Flavour Puzzle are often based on the use of symmetries
or else of special textures for the mass matrices. The fact that fermion mass matrices
related through weak basis transformations look different while leading to the same 
physics raises the question of how to recognise the same model written in different 
bases where the symmetry (or the special texture) is not apparent. A special technique
based on the use of weak basis (WB) invariants has been developed to 
tackle this problem. In this
talk special WB invariant conditions adapted to answer several different specific 
questions concerning leptonic physics are presented.  The same technique has been applied 
to the quark sector. The use of Higgs basis invariants based on the same rationale 
are extremely useful in the study of the scalar potential for multi-Higgs models.

\section{Building of weak basis invariant conditions}

The technique employed in this talk was developed for the fist time by the authors of
Ref.~\cite{Bernabeu:1986fc}  applied to the study of the  CP properties of the 
quark sector in the Standard Model. After spontaneous symmetry breaking quark masses
are generated and can be written as:
\begin{equation}
{\cal L}_m \mbox{(quarks)} = - \overline{u^0_L} \ m_u \ u^0_R -
 \overline{d^0_L} \ m_d \ d^0_R  +  \mbox{h.c.}
\label{qmas}
\end{equation}
still in a weak basis where, by definition of weak basis, the charged current is diagonal.
The charged current has the form:
\begin{equation}
{\cal L}_W \mbox{(quarks)}= - \frac{g}{\sqrt{2}}\   W^+_{\mu} \ 
 \overline{u^0_L} \ \gamma^{\mu} \ d^0_{L} +\mbox{h.c.}
\label{Wqua}
\end{equation}
Weak basis transformations are given by:
\begin{equation}
 d^0_L \longrightarrow U^\prime  d^0_L, \qquad
 u^0_L \longrightarrow U^\prime  u^0_L, \qquad
 d^0_R \longrightarrow W^\prime  d^0_R, \qquad
 u^0_R  \longrightarrow  V^\prime r^0_R
\label{WB}
\end{equation}
with $U^\prime$, $V^\prime$ $W^\prime$ arbitrary unitary matrices,
whereas the most general CP transformation for fermion fields in a weak basis allows 
for the combination of the definition of the CP transformation of a single fermion with 
a weak basis transformation \cite{Ecker:1981wv}, this is so because the fermionic CP 
transformation must be defined taking into account the part of the Lagrangian that conserves 
CP, to wit, the fermion gauge couplings. Pure gauge theories together with fermions do not 
violate CP \cite{Grimus:1995zi}. The most general CP transformation allowed by these 
couplings is then:
\begin{eqnarray}
{\rm CP} u^0_L ({\rm CP})^{\dagger}&=&U^\prime
\gamma^0  C~ \overline{u^0_L}^T;  \quad
{\rm CP} u^0_R({\rm CP})^{\dagger}=V^\prime \gamma^0  C~ \overline{u^0_R}^T
\nonumber \\
{\rm CP} d^0_L ({\rm CP})^{\dagger}&=&U^\prime \gamma^0 C~
\overline{d^0_L}^T; \quad
{\rm CP} d^0_R ({\rm CP})^{\dagger}=W^\prime \gamma^0  C~
\overline{d^0_R}^T \label{cp}
\end{eqnarray}
In order for the Lagrangian to be CP invariant the following relations have to be verified:
\begin{eqnarray}
U^{\prime \dagger} m_u V^\prime &=& {m_u}^* \label{cpmu} \\
U^{\prime \dagger} m_d W^\prime &=& {m_d}^* \label{cpmd} 
\end{eqnarray}
this means that there must be unitary matrices $U^\prime$, $V^\prime$ $W^\prime$ that 
obey these relations. For real mass matrices these relations are trivially satisfied 
with identity matrices. Looking for such unitary matrices in more general cases is not simple, 
however by combining these relations in such a way as to produce similarity transformations 
it is possible to obtain simple conditions
expressed only in terms of the mass matrices by applying traces and determinants. In this way
one may obtain a simple condition \cite{Bernabeu:1986fc}:
\begin{equation}
{\rm tr} \left[ H_u,   H_d \right]^3 = 0
\label{lll}
\end{equation}
which is a necessary and sufficient condition for CP conservation in the SM.
For three generations this condition is equivalent to 
\begin{equation}
{\rm det} \left[ H_u,   H_d \right] = 0
\label{trr}
\end{equation} 
which was obtained in Ref.~\cite{Jarlskog:1985ht} for the particular weak basis
where the quark mass matrices are Hermitian. However this choice of a particular basis
is not necessary since both conditions are weak basis invariant.

Ref.~\cite{Bernabeu:1986fc} was the starting point for the development of an
extremely powerful technique to test for CP violation in many different scenarios.

\subsection{Testing for CP violation, leptonic sector, low energies}
At low energies, assuming that the lepton number is violated, one can write an effective
Majorana neutrino mass matrix and the leptonic mass terms are then of the form:
\begin{equation}
{\mathcal{L}}_{\mbox{mass}} = -\frac{1}{2} {\nu_L^{0}}^T C^{-1} m_\nu
\nu_L^{0} - \overline{\ell_L^0 } m_\ell \ell_R^0 + \mbox{h.c.}\ ,
\end{equation}
the charged currents have a form similar to the one of Eq.~(\ref{Wqua}) written 
in terms of leptons. The WB transformations in the leptonic sector are given by:
\begin{equation}
\nu_L^{0} \rightarrow V \nu_L^{0}, \quad \ell_L^0 \rightarrow V \ell_L^0,
\quad \ell_R^0 \rightarrow W \ell_R^0
\end{equation}
with $V$ and $W$ unitary $3 \times 3$ matrices. 
Leptonic mixing and CP violation in the leptonic sector is described by the 
Pontecorvo-Maki-Nakagawa-Sakata (PMNS), $U_{PMNS}$ matrix. Using the standard 
parametrisation \cite{Olive:2016xmw} this matrix can be written:
\begin{eqnarray}
U_{PMNS} =\left(
\begin{array}{ccc}
c_{12} c_{13} & s_{12} c_{13} & s_{13} e^{-i \delta}  \\
-s_{12} c_{23} - c_{12} s_{23} s_{13}   e^{i \delta}
& \quad c_{12} c_{23}  - s_{12} s_{23}  s_{13} e^{i \delta} \quad 
& s_{23} c_{13}  \\
s_{12} s_{23} - c_{12} c_{23} s_{13} e^{i \delta}
& -c_{12} s_{23} - s_{12} c_{23} s_{13} e^{i \delta}
& c_{23} c_{13} 
\end{array}\right) \cdot  P  \label{std}  \nonumber \\
P=\mathrm{diag} \ (1,e^{i\alpha_{21}}, e^{i\alpha_{31}}) \qquad \qquad \qquad \qquad
\qquad \qquad \qquad \qquad
\end{eqnarray}
where $c_{ij} \equiv \cos \theta_{ij}$ and $s_{ij} \equiv \sin \theta_{ij}$.
There are three CP violating phases, $\delta$, $\alpha_{21}$, and $\alpha_{31}$
in Eq.~(\ref{std}).  In Ref.~\cite{Branco:1986gr}
a set of necessary and sufficient WB invariant conditions for CP invariance 
were derived, valid in the case of three generations for nonzero and nondegenerate 
masses:
\begin{eqnarray}
\mbox{Im}\  \mbox{Tr} \ [h_{\ell} \cdot m_{\nu}^*\cdot m_{\nu} \cdot m_{\nu}^* \cdot h_{\ell}^* 
\cdot m_{\nu} ]  
\  = 0  \label{asd} \\
\mbox{Im}\  \mbox{Tr} \ [ h_{\ell} \cdot( m_{\nu}^*\cdot m_{\nu})^2 \cdot (m_{\nu}^* 
\cdot h_{\ell}^* \cdot m_{\nu}) ] = 0\\
\mbox{Im}\  \mbox{Tr} [ h_{\ell} \cdot( m_{\nu}^*\cdot m_{\nu})^2 \cdot (m_{\nu}^* 
\cdot h_{\ell}^* \cdot m_{\nu})(m_{\nu}^* \cdot m_{\nu}] = 0\\
\mbox{Im}\  \mbox{Tr} [(m_{\nu} \cdot h_{\ell} \cdot m_{\nu}^*) + (h_{\ell}^* \cdot m_{\nu}
\cdot m_{\nu}^*)] = 0
\end{eqnarray}
In Ref.~ \cite{Dreiner:2007yz} a minimal set of necessary and sufficient conditions
for CP invariance was given:
\begin{eqnarray}
\mbox{Tr}\  [m_{\nu}^* \cdot  m_{\nu}^T, \ h_\ell]^3 = 0   \label{13} \\
\mbox{Tr}\  [m_{\nu} \cdot h_\ell \cdot  m_{\nu}^*, h_\ell^*]^3 = 0 \label{14} \\
\mbox{Im} \mbox{Tr} \ (h_{\ell} \cdot m_{\nu}^*\cdot m_{\nu} \cdot m_{\nu}^* 
\cdot h_{\ell}^* \cdot m_{\nu}) = 0 \label{15}
\end{eqnarray}
Eq.~(\ref{13}) is similar to Eq.~(\ref{lll}) which was derived for the quark sector,
this invariant is only sensitive to the Dirac-type phase, $\delta$.
The other two invariants are sensitive both to Dirac and Majorana type phases.
The invariant of Eq.~(\ref{14}) was first derived in Ref~\cite{Branco:1998bw} 
applied to the study of CP violation in the context of three degenerate neutrinos
which still allows for Majorana-type CP violation.
The invariant of Eq.~(\ref{15}) coincides with Eq.~(\ref{asd}) and was derived before
in Ref.~\cite{Branco:1986gr}, applied to the case of two generations, which
also allows for Majorana-type CP violation.

\subsection{Testing for CP violation, leptonic sector, Leptogenesis}
In the minimal seesaw framework \cite{Minkowski:1977sc,Yanagida:1979as,Levy:1980ws,GellMann:1980vs,Mohapatra:1979ia},
one introduces righthand neutrino fields which are 
singlets of  $SU(2)\times U(1)$. The most general leptonic mass term may then be 
written as:
\begin{equation}
{\cal L}_m  = -[\overline{{\nu}_{L}^0} m_D \nu_{R}^0 +
\frac{1}{2} \nu_{R}^{0T} C M_R \nu_{R}^0+
\overline{l_L^0} m_l l_R^0] + h. c.
\end{equation}
Let us assume that three righthanded neutrinos are introduced, this does not need 
to be the case, one needs at least two. The scale of $M_R$ can be much higher than the 
electroweak scale and in this case the seesaw mechanism operates. In the context of
seesaw, it is possible to generate a lepton number asymmetry through the decays of
the heavy Majorana neutrinos. This is the so-called  Leptogenesis mechanism 
\cite{Fukugita:1986hr,Davidson:2008bu,Branco:2011zb} and requires
CP violation at high energies. In the general seesaw framework it is not possible 
to establish a connection between leptonic CP violation 
at low energies and CP violation at high energies \cite{Branco:2001pq,Rebelo:2002wj}. 
Such a relation can only be established  in the context of a flavour theory. 

For Leptogenesis in the single flavour approximation, i.e., in the case when
washout effects are not sensitive to the different flavours of charged leptons into 
which the heavy neutrino decays, the possibility of having Leptogenesis can be probed 
by means of the following invariant conditions \cite{Branco:2001pq}:
\begin{eqnarray}
I_1 \equiv {\rm Im Tr}[m_D m^\dagger_D M^\dagger_R M_R  M^*_R m^T_D m^*_D M_R]=0 \label{i1} \\
I_2 \equiv {\rm Im Tr}[m_D m^\dagger_D (M^\dagger_R M_R)^2 M^*_R m^T_D m^*_D M_R] = 0 
\label{i2} \\
I_3 \equiv {\rm Im Tr}[m_D m^\dagger_D (M^\dagger_R M_R)^2 M^*_R m^T_D m^*_D M_R M^\dagger_R M_R ] = 0 \label{i3}
\end{eqnarray}
The first one of these invariants was discussed in Ref.~\cite{Pilaftsis:1997jf}.
For a detailed discussion of these invariant conditions see Ref.~\cite{Branco:2001pq}.
In the case of flavoured Leptogenesis additional CP-odd weak basis 
invariant conditions are required. A simple choice of additional invariant 
conditions is obtained by replacing $m_D m^\dagger_D$ in 
the above equations by $m_D h_l m^\dagger_D$, where 
$h_l = m_l m^\dagger_l$ \cite{Branco:2001pq}. 

\subsection{Beyond tests for CP violation. Texture zeros}
The imposition of texture zeros in the Yukawa couplings allows to establish a connection
between low energy physics and physics at high energies, for instance Leptogenesis. 
In Ref.~\cite{Branco:2005jr} we addressed the question of how to recognise flavour models 
corresponding to a set of texture zeros when written in a different weak 
basis where the zeros are not explicitly present. We considered texture zeros 
in $m_D$ in the seesaw framework, appearing in the weak basis where $M_R$ and $m_l$
are diagonal and real and we found invariants that vanish for different classes of textures.
One such example is:
\begin{equation}
I= {\rm Tr} [m_DM^\dagger_RM_Rm^\dagger_D, h_l]^3 
\end{equation}
Implications for two texture zeros in $m_D$ in the case of two righthanded neutrinos 
were studied in \cite{Ibarra:2003up}.
In Ref~\cite{Branco:2007nb} we classified all allowed four zero textures in $m_D$ with three righthanded neutrinos
and we showed that in general CP may be violated both at low and high energies. Furthermore,
in all these cases the parameters relevant for Leptogenesis can be fully specified in terms of
light and heavy neutrino masses and low energy leptonic mixing.

\subsection{Beyond tests for CP violation. The octant of $\theta_{23}$ together with the 
neutrino mass ordering}
In Ref.~\cite{Branco:2017koj} we have built weak basis invariants that provide a
clear indication of whether a particular lepton flavour model
leads to normal or inverted hierarchy for neutrino masses and what is the octant of 
$\theta_{23}$, of the standard parametrisation of $U_{PMNS}$.
It was shown in Ref.~\cite{Branco:2017koj} that the sign of the invariant:
\begin{equation}
\tilde{I_{1}}\equiv Tr[H_{\ell }\ H_{\nu }]-\frac{1}{3}Tr[H_{\ell
}]Tr[H_{\nu }]
\end{equation}
indicates the ordering of the neutrino masses and that the invariant:
\begin{equation}
\tilde{I_{2}}\equiv Tr[H_{\ell }]\ Tr[H_{\ell }^{2}\ H_{\nu }]-Tr[H_{\ell
}^{2}]Tr[H_{\ell }\ H_{\nu }]  \label{i22}
\end{equation}
is sensitive to the $\theta_{23}$ octant. In Table1 we show how to combine the information
provided by these two invariants.
\begin{table}[htb]
\caption{Combination of the two invariants. NO stands for normal 
ordering, IO for inverted ordering}
\label{Table:combination}
\begin{center}
\begin{tabular}{|c|c|c|}
\cline{2-3}
\multicolumn{1}{c|}{} &  &  \\
\multicolumn{1}{c|}{} & $\tilde{I_{2}} > 0$ & $\tilde{I_{2}} < 0$ \\
\hline
  & &   \\
$\tilde{I_{1}} > 0$ & NO, $\theta_{23} < \pi/4$ & 
NO, $\theta_{23} > \pi/4$\\
\hline
  & &   \\
$\tilde{I_{1}} < 0$ & IO, $\theta_{23} > \pi/4$ & 
IO, $\theta_{23} < \pi/4$\\ 
\hline 
\hline
\end{tabular}
\end{center}
\end{table}
In Ref.~\cite{Branco:2017koj} these invariants were applied to specific Ans\" atze studied in 
the literature\cite{Frampton:2002yf}. For a different strategy see \cite{Cebola:2015dwa}.

\section{Conclusions}
This talk presents a very powerful tool based on the derivation of weak basis invariant conditions
that are extremely useful for model building. Such conditions have been widely used and derived in 
the literature in different contexts by many different authors. Ref.~\cite{Branco:1999fs} provides a 
long list of references for conditions relevant to the study of CP violation both in the quark and in the 
leptonic sector, as well as in the Higgs sector, in several extensions of the Standard Model.

\section*{Acknowledgments}
The author thanks the Organisers of the 18th Lomonosov Conference
for the very fruitful scientific meeting and the warm hospitality.
This work was partially supported by Funda\c{c}\~ao
para a Ci\^encia e a Tecnologia (FCT, Portugal)
through the projects CERN/FIS-NUC/0010/2015, and
CFTP-FCT Unit 777 (UID/FIS/00777/2013) which are partially
funded through POCTI (FEDER), COMPETE, QREN and EU.


\end{document}